\documentclass[11pt]{report}
\usepackage{amsfonts}
\usepackage[final]{pdfpages}
\usepackage{ucs}
\usepackage{color,graphicx}
\usepackage{amssymb}
\usepackage{wrapfig}
\usepackage{setspace}
\usepackage{version}
\usepackage{fullpage}
\usepackage{framed}
\usepackage[color]{changebar}
\usepackage{program}
\usepackage[numbers]{natbib}
\usepackage{graphicx}
\usepackage{caption}
\usepackage{subcaption}

\cbcolor{blue}

\begin{document}

\begin{flushleft}

\Large

{\bf Mathematical modeling of the metastatic process.}

\normalsize

Jacob G.\ Scott$^{1,2}$, Philip Gerlee$^{3,4}$, David Basanta$^1$, Alexander G.\ Fletcher$^2$, Philip K.\ Maini$^2$ \& Alexander R.A.\ Anderson$^1$\\

$^1$ Integrated Mathematical Oncology, H. Lee Moffitt Cancer Center and Research Institute, Tampa, FL, USA\\
$^2$ Centre for Mathematical Biology, Mathematical Institute, Oxford University, UK\\
$^3$ Sahlgrenska Cancer Center, University of Gothenburg; Box 425, SE-41530 Gothenburg, Sweden \\
$^4$ Mathematical Sciences, University of Gothenburg and Chalmers University of Technology, SE-41296 Gothenburg, Sweden \\

\vfill
Correspondence:\\
Jacob G. Scott, jacob.g.scott@gmail.com

Running title: Mathematical models of metastasis

\end{flushleft}

\newpage

\section*{Abstract}
Mathematical modeling in cancer has been growing in popularity and impact since its inception in 1932. The first theoretical mathematical modeling in cancer research was focused on understanding tumor growth laws and has grown to include the competition between healthy and normal tissue, carcinogenesis, therapy and metastasis.  It is the latter topic, metastasis, on which we will focus this short review, specifically discussing various computational and mathematical models of different portions of the metastatic process, including: the emergence of the metastatic phenotype, the timing and size distribution of metastases, the factors that influence the dormancy of micrometastases and patterns of spread from a given primary tumor.

\newpage

\section*{Introduction: Why use mathematical models?}

Metastasis accounts for 90\% of cancer related deaths \cite{Weigelt:2005fk}, and the shift from localized to metastatic disease represents a paradigm shift for clinicians and patients alike as the strategy for therapy changes from aggressive and localized, to systemic and generally palliative. Despite its importance, this complex multi-step process remains poorly understood.  With the exception of studies showing genetic correlations between primary sites and sites of metastatic arrest \cite{Bos:2009jl,Minn:2005ia}, there is little understanding of the driving principles behind this process. Our lack of knowledge is for example, reflected in the fact that self-seeding, a process whereby a primary tumor releases metastatic cells that return to the primary tumor and accelerate its growth, is hypothesized to be a driver of primary growth \citep{Norton:2006us}, yet our current knowledge of the metastatic cascade is insufficient to determine the validity of this claim. For such a multi-faceted process, only through a combination of experimental and theoretical investigations can we hope to gain a comprehensive mechanistic understanding, and therefore uncover sensitive points where we can intervene and prolong the life of affected patients.

\begin{figure}[!htb]
\begin{center}
\includegraphics[scale=0.5]{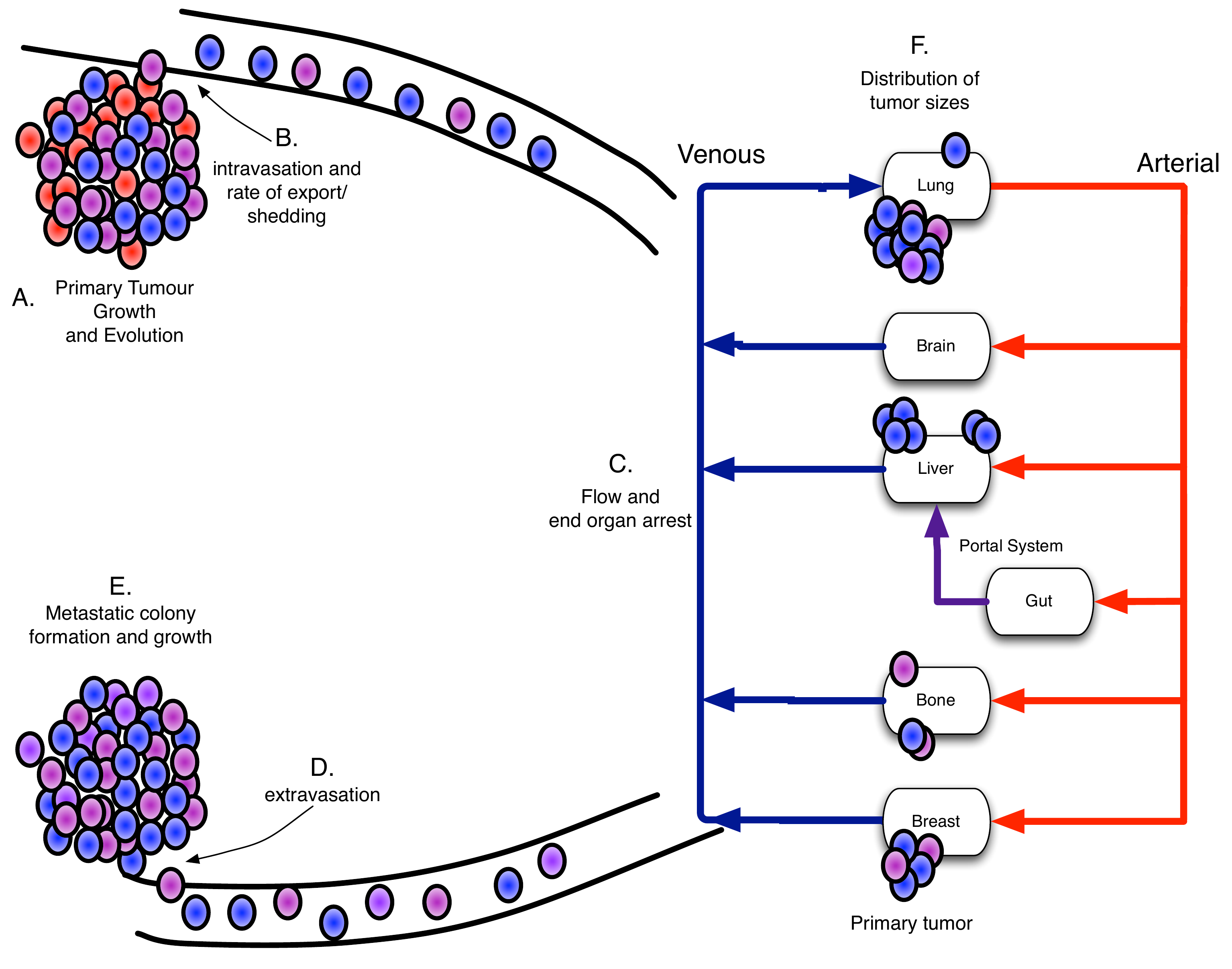}
\caption{\label{fig:overviewfig}
An overview of the metastatic process.  Each step in the cascade represents an opportunity for experimental systems to be designed.  Understanding the temporal dynamics within each step, and how the steps join together, however, is a challenge that also requires mathematical modeling. Cells from typically heterogeneous primary tumors grow (A) and some (blue and purple) are able to intravasate (B). Once in the vascular system, cells are subject to physical forces and selection of flow and filtration (C) until they extravasate (D) and colonise a foreign tissue bed (E).  tumors at this final stage will be distributed in size based on temporal and other factors (F) and will be made up of only certain clones from the primary tumor dependent on biological factors.
}
\end{center}
\end{figure}

The strength of experimental model systems is their ability to provide clear answers to specific questions. The strength of mathematical models is their ability to combine disparate experimental data and coalesce them into a coherent framework, which can then be used to predict the overall dynamics of the system in question. In particular, mathematical models allow for identification of the parameters to which the system is the most sensitive, and also allow for logical reasoning beyond what experiments can provide. In this sense, mathematical models of metastasis should be playing a larger role in the research in this area as experimentation is typically limited to one or a few steps in the cascade.  Many mathematical models to date, however, have also concentrated on only a subset of the steps of the cascade.  While these models are useful, as they have quantified the impact of parameters within the models of each step, they have yet to yield any fundamental additions to our knowledge of the process itself.

Mathematical models of biological systems tend to fall into two broad camps: descriptive and mechanistic. Both approaches can provide useful insights but since we want to connect directly with experimental measurement and drive novel experimentation, mechanistic models are where we need to focus our attention. This doesn't mean that we should build all encompassing mechanistic models of every process that we think is important in metastasis - since this would only provide us with a complex caricature of the real system with no additional understanding. Instead, we need to consider key processes and describe them in a level sufficient to gain insight, which should be tied to the resolution of the experimental data that might be used to drive and validate such models in the first place. Mirroring experimental observation should always be a key part of model validation but ultimately if a model is to be useful it should also make predictions that go beyond current observations, and further drive our understanding and inquiry.

To this end, our group, and others, have begun to build models more like those typically built to understand complex engineering systems, into which the more detailed models can eventually be embedded when the time comes for specific predictions to be made.  Until that time, these higher level models serve to shed light on areas of our knowledge which are most severely lacking, and provide experimental questions to fill those gaps in a systematic manner.

\begin{figure}[!htb]
\begin{center}
\includegraphics[scale=0.5]{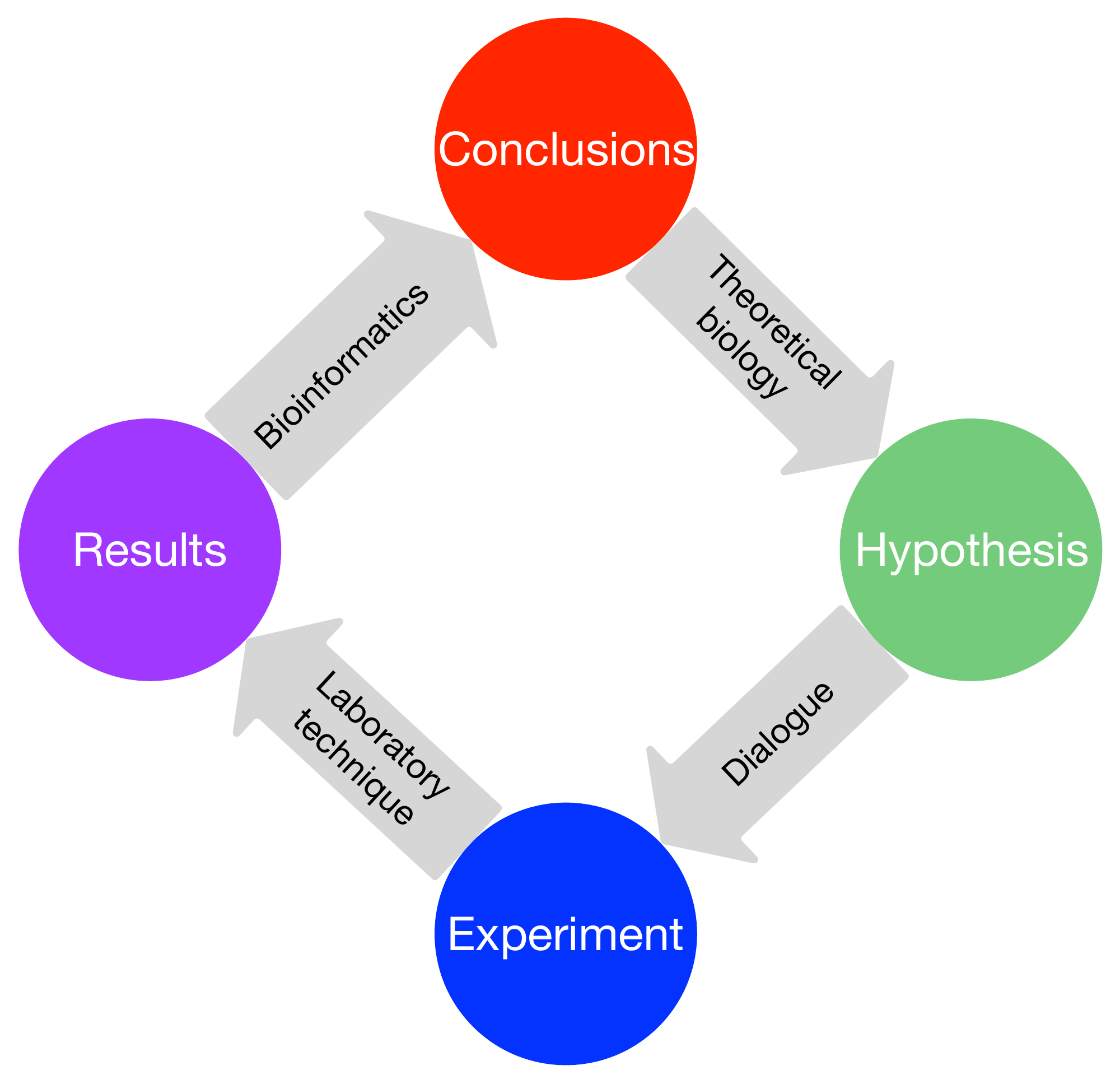}
\caption{\label{fig:scientificmethod}
An overview of the scientific method, and where theoretical/computational scientists fit into this process in the life/medical sciences and biology. Theoretical biology is the science of putting together existing knowledge into specific theoretical frameworks which can be used to make predictions and generate further hypotheses.  Bioinformatics is a statistical science used for helping scientists make conclusions when faced with large data sets and non-linear relationships. Dialogue between multi-disciplinary scientists helps shape meaningful experiments.  Laboratory technique translates experimental constructs into meaningful results.
}
\end{center}
\end{figure}

\section*{Where experiments cannot go: opportunities for mathematical models}

To definitively answer questions in biology, the burden of proof falls on the experimental scientists.  To connect the disparate experimental `truths' into a coherent framework however, is the purview of theorists.  Further, when theory has been established, but necessary experimental techniques have yet to be developed, theory can again step in to advance science by making specifically testable predictions - vastly shrinking the set of possible experiments.  This dual role is best understood by thinking of where mathematical/computational science fits into the scientific method.  The place of theory, typically called mathematical or theoretical biology, lies between biological conclusions and new hypotheses (Figure \ref{fig:scientificmethod}).  The central goal of this discipline is to create rigorous frameworks, beyond linear `cartoon' models of mechanism, through which specific predictions and hypotheses can be generated.  Bioinformatics, another computational field in biology, largely works between experiment and conclusion, helping to make sense of the vast amounts of data that come out of modern day experiments.  This review will focus on models from the field of theoretical biology.

\subsection*{A note on mathematical tools: stochastic vs. deterministic models}

Like experimentalists, mathematical modelers have many tools at their disposal.  These tools comprise a wide spectrum, ranging from classical `pen and paper' models, to those requiring powerful computers to iterate, and everything in between.  There are many ways to distinguish these models from each other, but likely the most telling dichotomy is the difference between stochastic and deterministic models.  

A deterministic model is one in which there is no randomness: the model will behave exactly the same way each time it is solved.  This does not mean that the model is necessarily predictable: indeed, many deterministic models exhibit wild fluctuations and even chaotic results, exhibiting strong dependence on even small changes in initial conditions or parameter values.  The strength of these models is that they allow us to understand all the possible behaviors of a system and in which parameter regimes those behaviors occur. Examples of deterministic models discussed in this review include ordinary and partial differential equations (ODEs and PDEs) which describe how key quantities of interest, such as chemical concentrations or cell densities, vary continuously with one (in the case of ODEs) or many (for PDEs) independent variables.  Well mixed systems, where space is not considered, are typically modeled with ODEs where the dependent variables evolve in time; PDEs are utilized when there is also spatial heterogeneity, or differences in `age' or differentiation status across the population modeled.

A stochastic model, on the other hand, has randomness written in to the system itself. This randomness can be incorporated into the model in many different ways.  Models can represent many individual entities which can interact with one another or move in ways defined by probabilities. Alternatively, noise terms may be explicitly incorporated into existing deterministic descriptions.  As compared to their deterministic counterparts, these sorts of models often are better representations of the underlying biological processes, which do seem to be governed at some level by randomness, and the results of any given simulation of a model can be quite different from another, again mimicking biology.  Gaining a deep understanding of these models through analysis however, is usually much more difficult, and we must often rely upon averages of many realisations to gain an understanding of the system, or on analysing the average ("mean-field") behavior of the system, effectively returning to a deterministic description.  Examples of the types of stochastic model discussed in this review include Markov chains (MCs), cellular automata (CAs) and Moran processes. MCs are stochastic processes in which a population is subject to a time-independent series of transitions, from one state to another in a `memoryless' fashion, that is without regard to the history of the system \cite{Norris:1998}.  CAs are a discrete time and discrete `cell' based models. Individual cells, often called `agents', are programmed with simple rules and simulated as they interact in a computational domain.  Complex behaviors often can emerge from simple rule sets and limited numbers of agents \cite{vonNeumann:1966}, and while these models are not necessarily stochastic by nature, the ones described in this review are, by virtue of their rules. The Moran process \cite{Moran:1962uq} is a stochastic process of birth and death in which a mixed population of two (or more) species compete in a manner meant to mimic Darwinian selection.

In the science of metastasis, all of these model types have been, and will continue to be, utilized for different applications.  This is particularly relevant as there are a number of steps in the metastatic cascade (outlined in Figure \ref{fig:overviewfig}) which span multiple scales, both spatial and temporal.  In the remainder of this brief review, we will cover the most relevant mathematical models of the metastatic cascade and highlight the ways in which these models have affected our knowledge, future experiments and clinical decision making.  We will begin by describing a series of mathematical models built around a specific experimental murine metastasis model which was able to give insight into several key, unmeasurable parameters.  We will then describe a series of models focussed on the genetic emergence of metastasis which generates a number of hypotheses, but can not yet be tied to experimental data. We then discuss models aimed at understanding the size distribution of metastasis at the time of diagnosis, several of which offer the possibility of connecting to patient-specific data. We then review several models aimed at understanding the temporal patterns of recurrence through the dormancy mechanism, and we conclude by reviewing some recent network models of metastasis aimed at understanding the anatomic patterns of metastatic spread within the body.

\subsection*{Models of experimental systems} 

One of the first attempts to model the metastatic process is described in a series of papers by Liotta and colleagues \cite{Saidel:1976tl,Liotta:1976tx,Liotta:1977us}.  In this work, the authors built an experimental system and mathematical models in parallel in an attempt to better define the parameters of each step in the process. The experimental system considered was a mouse model of fibrosarcoma that readily formed pulmonary metastases,  via both implantation and intravenous injection. In this way, the authors were able to accurately control many parameters, and use the results to obtain estimates for those that could not be measured directly. 

\begin{figure}[!htb]
\begin{center}
\includegraphics[scale=0.5]{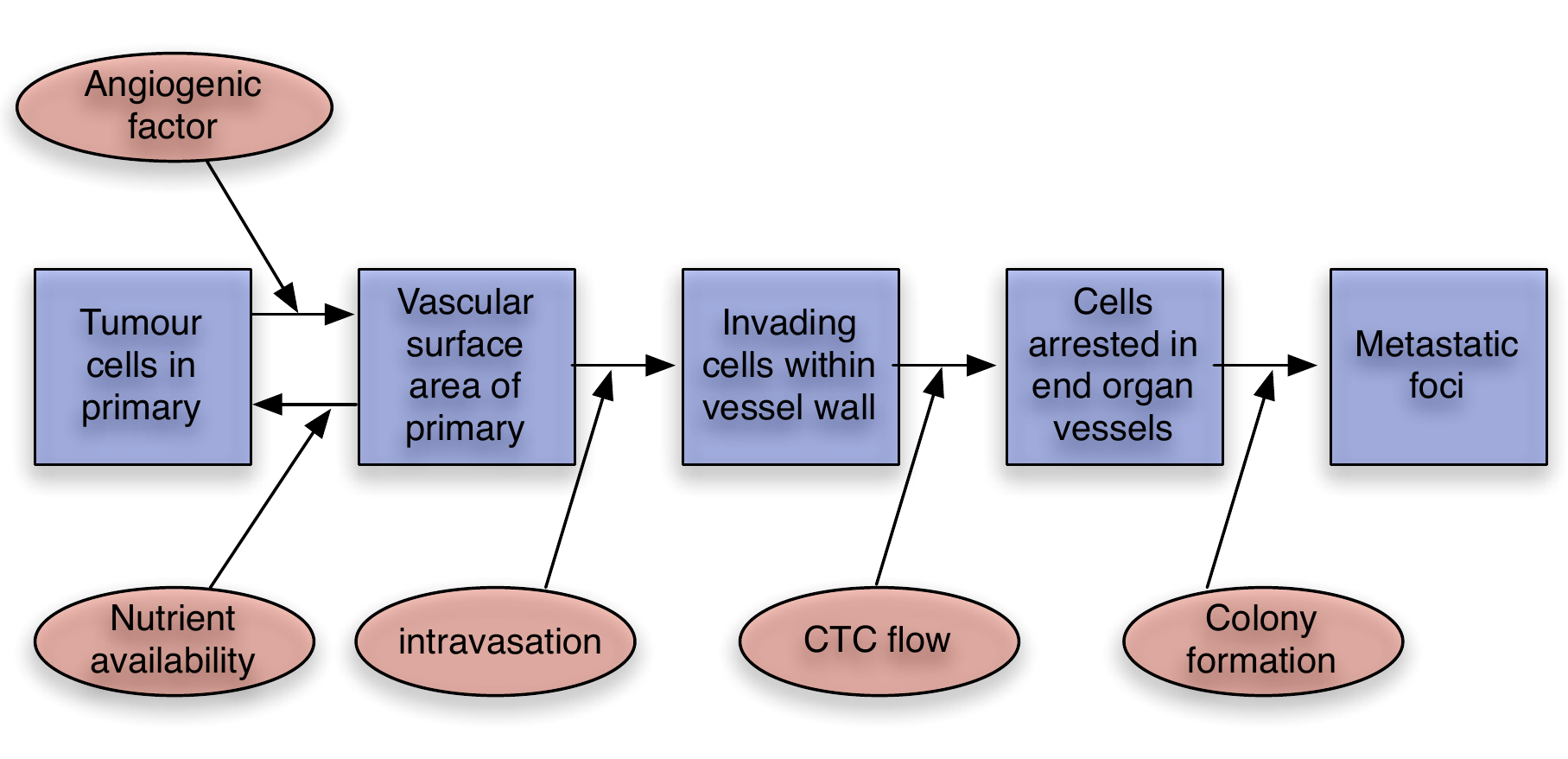}
\caption{\label{fig:liottafig}
Compartment model developed by Saidel et al. \cite{Saidel:1976tl}  Each of the blue boxes represent measurable quantities from the \textit{in vivo} model, while the ovals represent quantities inferred from the model.
}
\end{center}
\end{figure}

The authors derived an ODE model to describe how the population of tumor cells changes in time in each of several key `compartments' as a result of flux between them.  Each of the compartments in their mathematical model represents a discrete phase of the metastatic cascade, illustrated schematically by the blue boxes in Figure  (\ref{fig:liottafig}) \cite{Saidel:1976tl}.  Numbers of cells measured from the murine experimental system were used to parameterize this mathematical model.  The model was then used to predict the effect of a number of perturbations to the system including: tumor resection, tumor trauma, vessel growth inhibition, lung vessel damage and inhibition of intravasation.  Each of these perturbations was simulated to predict response and the experimental system was then assayed with good correlation, giving significant insight into the otherwise unmeasurable aspects of the system and the mechanisms driving the response to the perturbations.

As discussed previously, biological systems are often ruled by stochastic processes at the cellular level, and in the case of metastasis, this is certainly the case.  To this end, Liotta et al. developed and analysed a MC model of a subset of the above system \cite{Liotta:1976tx}. The authors additionally considered that cancer cells do not only travel and arrest as single cells, but are found in clumps of varying size \cite{Liotta:1974ug}. The model contains three compartments: tumor cell clumps in the circulation, tumor cell clumps arrested in the pulmonary capillary bed, and pulmonary metastatic foci. In agreement with experimental data, the model assumes that the entry rate of clumps is size-dependent, following a decaying power-law (i.e. the number of clumps of size $n$ scales as $n^{-\alpha}$, $\alpha = 2$), that the clump death rate is inversely proportional to size, and the colonisation rate increases linearly with clump size. The validation was carried out in the fibrosarcoma mouse model, where cancer cells were injected intravenously and the animals were sacrificed 10-30 days post-implant. The model showed good agreement both with respect to the number of macroscopic metastatic foci as a function of time, and the time-dependent probability of finding a metastasis-free animal. An interesting conclusion from the study was that larger clumps have a strong impact on metastasis formation, and hence that disassociating agents that reduce clump size could have therapeutic effect.

A follow-up study further simplified this mathematical model to account only for arrest and foci formation, without regard to clump size \cite{Liotta:1977us}. Interestingly, the sharp transition in the metastasis-free probability remained, suggesting that this phenomenon is due only to the stochastic nature of arrest and foci formation. If these results can be extended to patients they could represent a novel method for assessing the likelihood of micrometastatic lesions that eventually could become clinically relevant. 

\section*{Models as abstractions: insights into unmeasurable processes}

In the previously discussed models, each portion of the model corresponded directly to an aspect of an experimental system.  In this way, the authors were able to use the strength of each system to gain a deeper understanding of the mechanisms driving each portion of the well-controlled process. Many aspects of the metastatic process in the clinic, however, are not amenable to this sort of methodology, and cannot be measured/quantified directly.  This situation, where measurements are not yet able to be made, is one where mathematical models can play an influential role and relieve the impasse at which we would otherwise be.  A specific case of this is in tumor genomics: we know that genetic mutations play a role, but we are not yet able to measure the dynamic changes of a tumor genome within a patient over time.

\subsection*{Evolutionary models: emergence of metastatic clones}

The emergence of metastatic disease has largely been attributed to cells gaining functions specific to intravasation (Figure  \ref{fig:overviewfig} B).  This gain of function has been linked to genetic mutation, with large numbers of specific genes being implicated.  More generally, the epithelial-mesenchymal transition (EMT) has been identified as a process (likely polygenic) involved in the acquisition of metastatic potential \cite{Chaffer:2011fk}.

Experimental studies have shown that EMT (among other phenotypic changes) is important for the development of metastatic clones \cite{Thiery:2003uq}, but as measuring the individual mutations within a patient's tumor over time remains beyond the scope of experimental science, understanding the dynamics of this process is a ripe question for theoreticians.  To this end, a number of models have employed a stochastic description called the Moran process \cite{Moran:1962uq} to study the genetic landscape of a tumor's cellular population over time.  In this process, populations of constant size consisting of individual agents (cells), usually of two distinct (geno)types, are grown in competition with one another under selection.  The Moran process was originally designed to mimic Darwinian selection, where cells of a given type are chosen randomly to divide or die based on an ascribed fitness, usually linked to division rate.  The population dynamics are then simulated, with the aim of understanding long-term behavior (coexistence or dominance by one population).  These models serve as excellent platforms through which to understand the emergence of new clones within a population, much like the emergence of EMT, or any other metastasis-specific trait.

\begin{figure}[!htb]
\begin{center}
\includegraphics[scale=0.5]{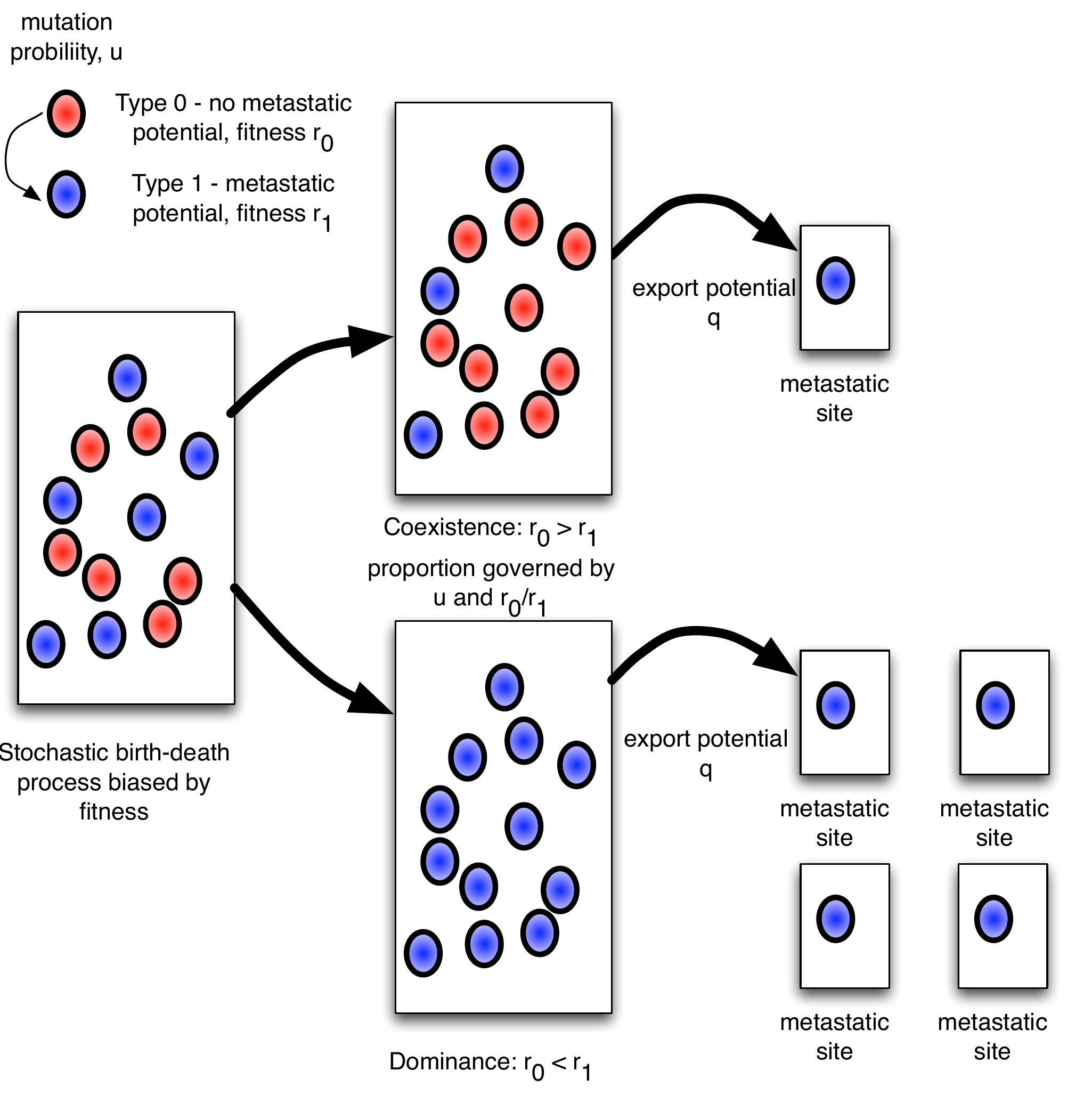}
\caption{\label{fig:moranoverview}
The Moran process as utilized to study the emergence of metastasis \cite{Michor:2006hl,Dingli:2007ut,Haeno:2010hd}. The primary tumor (left) is allowed to grow and turn over, the population changing based on the probability of mutation, $u$, and the relative fitness of the two cell types, $r_0$ and $r_1$. As cells gain the ability to metastasize (mutate into type 1 cells) they also have the opportunity, at rate $q$, to be exported and begin their own colonies. At steady state, the primary tumor can be composed of either all type 1 cells (dominance, when $r_1>r_0$) or a coexistence with the proportions of cells governed by the mutation $u$ and the fitness ratio ${r_1}/{r_0}$.
}
\end{center}
\end{figure}

To study the dynamics of the emergence of the metastatic phenotype, Michor et al. \cite{Michor:2006hl} proposed a model of tumor growth, based on the Moran process, that took account of mutation to a metastatic phenotype.  The authors modeled a heterogeneous tumor made up initially of cells without the ability to metastasize (type-0, fitness $r_0$). At each time step, a cell is randomly chosen to divide (biased by fitness) at which time the cell has a probability $u$ of producing mutated offspring that can metastasize (a type-1 cell) with fitness $r_1$ (where a fitness of 1 is neutral). This mutated offspring also now has a probability $q$ of being `exported' from the population to initiate a metastatic tumor of their own (Figure \ref{fig:moranoverview}).  Results for a range of parameter combinations were calculated both analytically and by exact stochastic simulation.  The authors found that initiating tumors with different parameter combinations could lead to qualitatively different outcomes: the model predicted that metastatic clones are most likely the result of advantageous mutations that will occupy the majority of the primary tumor. Indeed, for a mutation that confers metastatic potential and simultaneously a lower fitness in the primary tumor, there must be approximately a million-fold increase in metastatic potential for it to generate the same number of metastases in a patient.

Dingli et al. \cite{Dingli:2007ut} extended the previous model by Michor et al. \cite{Michor:2006hl} by allowing tumors to grow above a constant size, and incorporating a dependence on tumor size in the export probability.  The authors suggested that certain types of mutations confer a fitness advantage ($r_1>1$) and metastatic ability (e.g.\ mutations in RAS and MYC), and can dominate the tumor and seed many metastases; while other types of mutations (such as MSG) have a lower relative fitness as compared to non-mutants ($r_1<1$) and can therefore co-exist only in small populations and can even be `exported' entirely, depending on the export rate ($q$).  This insight provides an explanation for the situation in which there exists metastatic disease without evidence for cells with metastatic potential in the primary, or in the more extreme case where there is no evidence of a primary tumor at all \cite{Greco:2012kx}.

To consider this model in a more clinically grounded context, Haeno and colleagues \cite{Haeno:2010hd} studied a new metric, total tumor burden, which they tied to survival.  In this study, the aforementioned model was extended and the timing of interventions, which included surgery (removal of a fraction of type-0 and type-1 cells; those residing in the primary tumor) and chemotherapy (affecting birth and death rates for all cells) was incorporated.  The authors used threshold values for total tumor burden to correlate with time of diagnosis and patient death and investigated the effect of each of the therapies.  They found that, depending on how a simulated patient's tumor was situated in parameter space (the relative rates of acquisition of the metastatic phenotype, $u$, `export' of the metastatic cells to foreign stroma, $q$, and the birth-death balance ($r_0$,$r_1$) of each cell type), qualitatively different outcomes could be obtained from therapies given at different times.  While currently beyond our abilities to tie to clinical data, this model served to illustrate how such a technique could shed light on the metastatic processes in play for a patient, and potentially influence treatment choice. 

\subsection*{Metastatic colony size distribution}

At present there is a single designation in the standard clinical cancer staging system (the TNM system, which describes the primary \underline{T}umour, any positive lymph \underline{N}odes and any \underline{M}etastasis) to describe metastatic disease: either M0 for a patient with no observable metastasis, or M1 for a patient with \textit{any amount} of metastatic disease; yet there can be great variation in both size and location of metastases from one patient to another.  Historically, patients with any amount of metastatic disease have only been offered localized treatment at those metastatic sites if they caused a specific problem, but not with curative intent (with several specific exceptions, e.g. solitary brain metastasis in lung cancer).  This paradigm is beginning to change with the advent of the concept of `oligometastasis' describing the situation where a patient may have only a small number of metastases, a number worth treating.  While only a small number of trials have been conducted \cite{Milano:2009zr,Milano:2012ys}, this approach is gaining in popularity with the increased availability of highly targeted, minimally invasive therapeutic modalities such as stereotactic body radiation therapy.  The main problem confronting this movement, however, is our lack of understanding of which situations represent `oligometastasis'.  That is, which patient with one obvious metastatic lesion actually has many other, subclinical ones, and which does not?  To answer this question, a number of mathematical models have been developed in an attempt to understand the distribution of sizes of metastatic lesions in time.

One such study is that of Iwata et al. \cite{IWATA:2000gi}, in which computed tomography (CT) images of spatially separated colonies of hepatocellular carcinoma in a patient's liver were fit to a novel PDE model of colony size. In their system the population of tumors was modeled as a distribution of colony size over time. Each colony was assumed to grow by a saturating growth function (specifically Gompertzian growth, though any growth law could have been used \cite{Gerlee:2013fk}) and release metastatic cells at a rate proportional to the volume of the colony raised to some power, effectively representing the fractal dimension of the blood supply of the tumor \cite{Baish:1998vn,Baish:2000yq}.

Using three successive scans of the patient's tumor progression without therapy, and several after initiation of chemotherapy, the model parameters were fit and predictions about the pre-diagnosis time course could be made.  Further, and likely of greater import, predictions about subclinical metastatic burden at the time of diagnosis were made.  This sort of information, which is currently not available to clinicians, represents a class of personalized information about a patient's disease that does not rely on genomic information, and could be measured for any patient who already has scans taken during the course of standard therapy. This approach, of using scans which are `standard of care', is being utilized in primary glioblastoma and is approaching clinical trials \cite{Neal:2013ys}, but the model of Iwata et al. represents the only such attempt, to our knowledge, in metastatic disease.

The same question that was addressed with a deterministic model by Iwata et al. \cite{IWATA:2000gi} has been addressed using a number of stochastic modeling techniques.  Bortoszyinski et al. \cite{Bartoszynski:2001ly}, Hanin et al. \cite{Hanin:2006fs} and Xu et al. \cite{Xu:1998vy} used similar growth laws as discussed by Iwata et al. and then derived expressions called joint distribution functions, which predicted the probability of there being a given distribution of metastatic colony sizes at a given time.  The authors then each validated their models against a single patient's data.  The models, after fitting, were also able to predict several salient features about the patients' pre-diagnosis condition and the natural history of their disease.  


\subsection*{Understanding temporal recurrence patterns} 

\subsubsection*{Tumor dormancy}

The mechanisms and timing of distant recurrence of cancers after treatment of the primary tumor remain difficult to study in the clinic.  It is widely believed that most patients have sub-clinical micrometastatic disease at the time of diagnosis, the distribution of which we discussed in the previous section, but that only some of them will go on to develop overt metastasis.  The reasons for this are largely unknown even though there is a large literature \cite{Paez:2012zr}, both experimental and theoretical, surrounding the period of so-called `tumor dormancy'.

\begin{figure}[!ht]
\begin{center}
\includegraphics[scale=0.5]{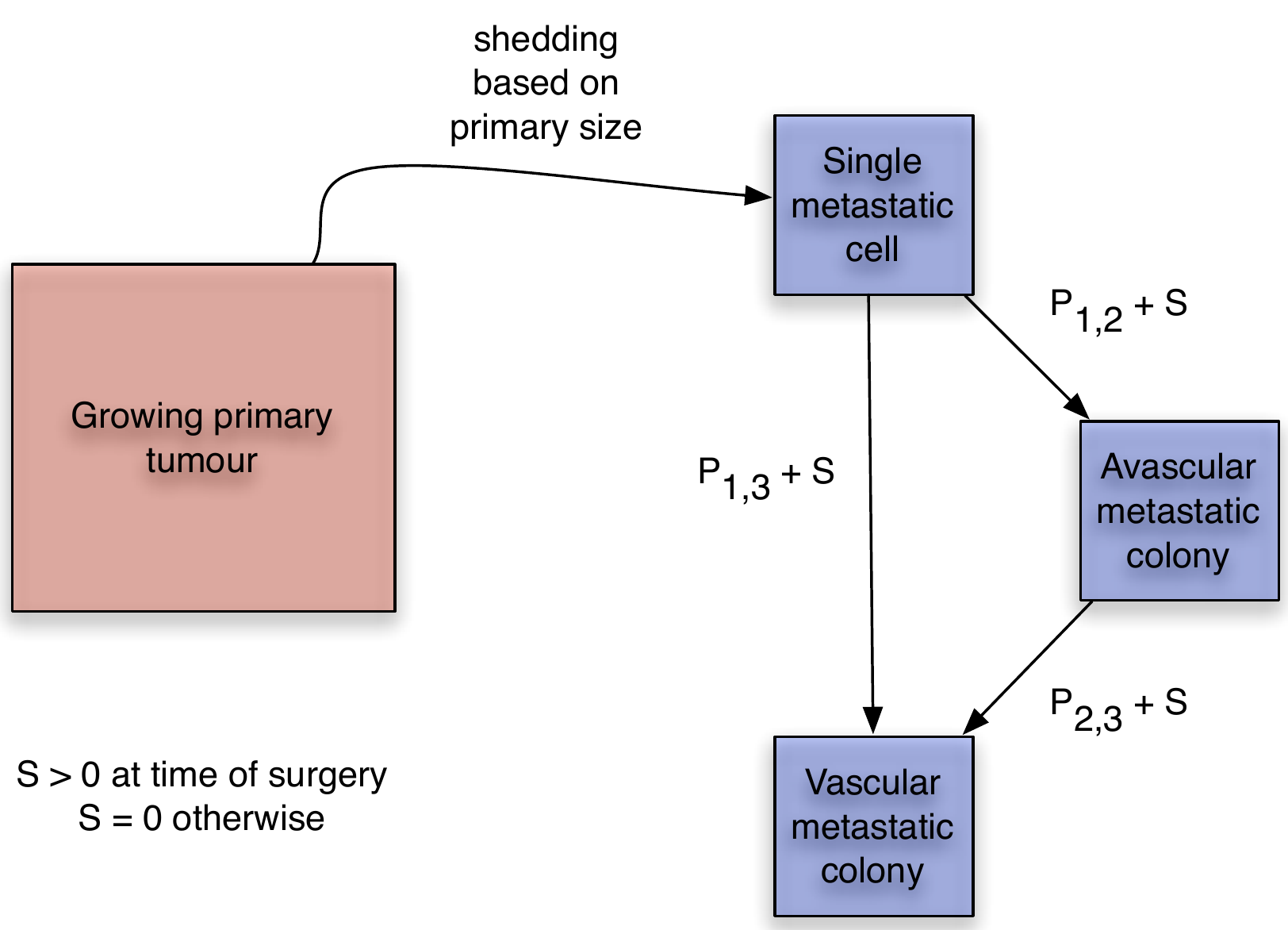}
\caption{\label{fig:retsky}
The model posited by Retsky et al. \cite{Retsky:1997vu}.  The primary tumor (red) is allowed to grow and randomly seed single metastatic cells.  The single cells can switch state stochastically to become growing colonies which are constrained by an avascular limit ($P_{1,2}$) or which are vascularized ($P_{1,3}$).  In order to fit the results of this model to bimodal recurrence pattern of the Milan trial \cite{Demicheli:1996uq}, they needed to effect a `bonus' to the transition probabilities at the time of surgery ($S$), which led to the hypothesis that there is a metastasis promoting role by peri-surgical inflammation.
}
\end{center}
\end{figure}

After definitive therapy for most primary cancers, the majority of distant recurrences occur in the first two years, mostly due to micrometastatic disease that was undetected at the time of primary therapy that eventually grew to a detectable size.  Demicheli and colleagues \cite{Demicheli:1996uq} however, noted a bimodal distribution of relapse times for patients treated in the Milan trial of primary surgery for breast cancer.  One peak was in the expected range, at 18 months, while the other was a broader peak centered at 60 months after surgery.  To understand this long lag time, Retsky, Demicheli and colleagues \cite{Retsky:1997vu,Demicheli:1997kx} proposed a new mechanism of cancer dormancy and recurrence.  They posited that micrometastases that exist at the time of surgery can be activated by the subsequent inflammation into a non-dormant state.  To illustrate their hypothesis they built a stochastic model of micrometastasis dormancy in which metastatic sites exist in one of three states: dormant single cells, colonies arrested at the avascular limit, and growing colonies.  In their simulations they allowed for stochastic transitions between these states (assuming the transition was to a larger state) and showed that this model could recapitulate the unexpected bimodal distribution of the large clinical trial - but only if they allowed for a transition `bonus' added at the time of surgery (Figure \ref{fig:retsky}).  Their results have been used to argue for differing chemotherapy schedules as well as suppression of inflammation at the time of primary surgery \cite{Demicheli:2007ys,Retsky:2012vn}.

Another explanation for variable dormancy times is related to the cancer stem cell hypothesis.  This well-known hypothesis has been modeled extensively for a number of different tumors (for a review, see Michor \cite{Michor:2008gu}), but has only received limited attention in connection to dormancy and metastasis.  Enderling and colleagues pioneered this work and showed, using a stochastic model of cellular hierarchy within a tumor, that single cancer stem cell-driven solid microtumors may undergo long periods of dormancy despite cellular activity \cite{Enderling:2009dk}.  The length of the dormancy period depends on the complex interplay between stem cells and their non-stem cancer cell counterparts.  Specifically, they found that impaired cancer stem cell migration, as well as large numbers of non-stem cancer cells, increase population dormancy times.  Higher non-stem cancer cell death rates were correlated with shorter dormancy times and, paradoxically, with increased tumor growth in the long term \cite{Enderling:2009dk,Enderling:2012cr,Enderling:2013dq}. This `tumor growth paradox' was also explored in an analytical model by Hillen et al. \cite{Hillen:2013oq} and was put forward as an explanation for some of the failures of therapy \cite{Gao:2013cz}.

\subsubsection*{Primary-secondary communication} 

The idea that primary tumor factors can affect the growth of metastases has also been modeled by considering communication between the primary tumor and secondary metastatic deposits.  Boushaba and colleagues \cite{Boushaba:2006zr,Kim:2013ly} considered an anti-angiogenic factor secreted by the primary which would keep spatially separated, yet local, metastases in a dormant state and reported a critical distance window in which this effect was active. This result is difficult to interpret in hematogenous metastasis as the idea of a diffusion `distance' for any factor secreted by the primary is not trivially understood because of the fluid dynamics involved in blood flow as compared to diffusion through tissue. A different study, by Eikenberry et al.  \cite{Eikenberry:2009qf}, considered the effect of the primary on metastatic deposits through interactions mediated by the immune system.  They modeled the removal of the primary as a decrease in immune stimulus, which in turn could promote metastatic growth.  A mechanism-agnostic model analyzed by Diego et al. showed primary-secondary communication to have an effect on metastatic growth, but only in a very small region of parameter space \cite{Diego:2012ve}, suggesting that, while possible, this is a rare phenomenon.  While there is a growing theoretical literature on this subject, the clinical data to support its import are lacking and the data in biological model systems have been shown in only a few studies, reviewed by Peeters et al. \cite{Peeters:2008bh}, and therefore it is difficult to draw any solid conclusions at this time.

\section*{Making sense of existing data}

We have now discussed mathematical models built with specific experimental systems in mind, ones designed to help explain some unmeasurable quantities in existing patients and ones for which no experiments can yet be done.  The final class of models that we will discuss were derived in order to analyze existing, population level data of metastatic spread, with the aim of making predictions about the most likely routes of spread. The aim of these models is not to examine and quantify the involved substeps (such as the models by Liotta et al. \cite{Saidel:1976tl,Liotta:1976tx,Liotta:1977us}), but instead they focus on the `global' system dynamics.

\subsection*{Patterns of spread: Metastasis dynamics on networks}

Understanding the patterns of spread of a particular primary tumor can help guide clinicians in their decision making for patients. This knowledge is useful for follow up purposes in that we can target our interrogations to the organs most likely at risk so as to minimize testing and maximize our chances of early detection of recurrence.  Further, understanding the temporal patterns of recurrence helps us to structure our follow up schedule and to understand when to employ the greatest vigilance, as early detection of recurrence gives the best chance of successful salvage therapy.  

This temporal aspect of metastatic spread was captured in a model by Chen et al.  \cite{Chen:2009hz}, which made use of a large database of Medicare claims. The data were such that, for each patient with a primary tumor, a temporal sequence of metastatic events labeled according to anatomical location were recorded. The authors analyzed the data by calculating a time-dependent hazard as a function of the primary and metastatic site, and could observe how, given a primary in a certain location, the risk of developing certain metastatic lesions developed over time. They also formulated a statistical model with the ability to predict the location of the primary tumor given a sequence of metastatic sites, and the reverse: given a certain primary, predicting the most typical sequence of metastatic sites.  The accuracy of the above predictions is, however, not yet of a quality that makes them a clinically relevant tool (the true positive rate of primary site prediction was 51 \%), although the study shows the potential of this kind of temporal data when mixed with a network based approach.

In another effort to better understand the patterns and timing of metastasis, Newton et al. built and analyzed a MC model of metastatic patterns of primary lung cancers \cite{Newton:2012bs,Newton:2013dg}.  By focusing on a specific cancer, rather than patterns overall, the authors hoped to be able to infer more about the mechanisms of metastasis than simply quantifying the patterns.  When building this model, they began by constructing a network of connected organs and made the assumption that any transition is possible as a direct step, that is: a cancer can move directly from any organ to any other organ (the network connectivity is `all to all'). Once this network was built, an iterative, random method called the Monte Carlo method \cite{Diaconis:2009} was used to solve for a series of transition probabilities, which would then lead to the steady state defined by a large autopsy study of untreated patients \cite{Disibio2008}.  The quantitative understanding provided by these studies goes beyond the empirical understanding clinicians have about patterns of spread from retrospective studies, and allows for a more detailed analysis of the parameters and the dynamics than is possible without these methods.  

The most important insight gained from this approach was that certain sites, in the case of primary lung cancer, act differently than others, and that these differences affect the metastatic patterns of the disease as a whole.  Specifically, Newton et al. \cite{Newton:2013dg} identify the adrenal gland and kidney as `spreaders', which, when colonized by metastases, significantly increase the probabiliy of further organs becoming involved.  They also identified regional lymph nodes, the liver and bone as `sponges', temporally suppressing metastasis in other sites when colonized.  


\subsection*{Embedding anatomically correct connectivity}

Each of the previous studies which has utilized a network-theoretic approach has assumed `all to all' connectivity.  In the case of hematogenous metastasis however, there is a simple and conserved network architecture (that of the vasculature) that significantly reduces the complexity of the problem and further, and more importantly, offers the possibility for patient specific modeling and prediction - something that the previous models are unable to do.  Specifically, the human vascular network can be written down very simply as a directed network which is weighted by relative blood flow and capillary bed filtration.  Scott and colleagues recently postulated a series of hypotheses based on this anatomically informed network \citep{Scott:2012vn}: that the specific filtration characteristics of each organ, modulated by the biology of the circulating tumor cells (CTCs), would significantly affect the half-life of CTCs in the circulation; that one could solve for metastatic patterns by knowing a patient's specific filtration characteristics, much like one could solve for quantities within an electric circuit; and that treatment could be personalized, based on CTC measurements from each of the vascular compartments, and knowledge of the primary tumor location.

\begin{figure}[!ht]
\begin{center}
\includegraphics[scale=0.5]{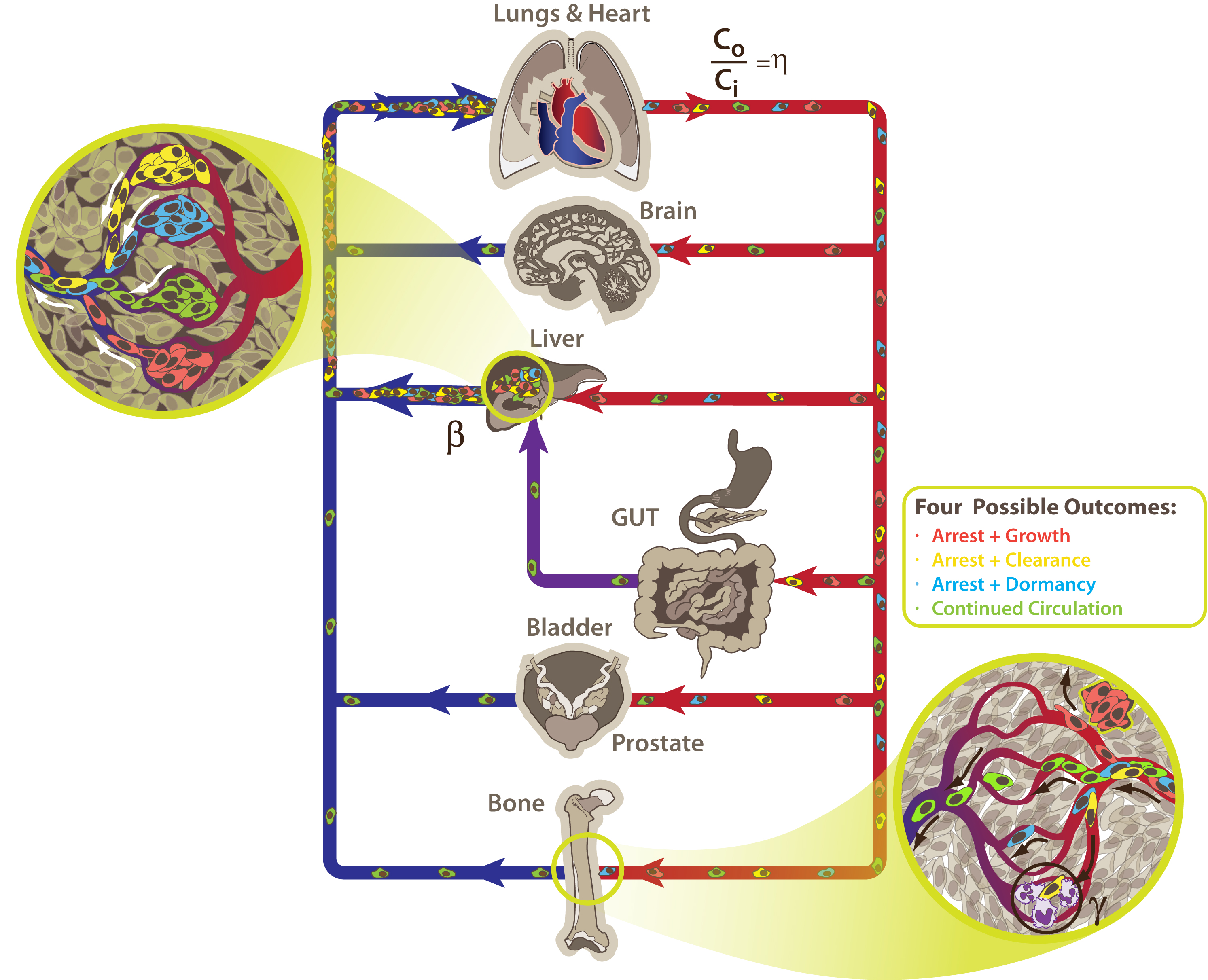}
\caption{\label{fig:filterflow}
The human vascular system represented as a network to illustrate the filter/flow perspective.  From this perspective, several new quantities can be calculated for an individual patient which could be used to tailor therapy in a personalized way.   Specifically: $\eta$, the filtration fraction, which is the proportion of CTCs that are able to traverse a given capillary bed; $\beta$, the shedding rate of an individual tumor; and, the number of CTCs in the each of the three distinct compartments (arterial, red; port venous, purple; and, systemic venous, blue). A knowledge of each of these could, for an individual patient, be used to better understand the individual's risk of metastatic spread.
}
\end{center}
\end{figure}

In a subsequent theoretical work, Scott et al. examined the self-seeding hypothesis \citep{Scott:2013hs} and showed that direct self-seeding (i.e. the primary tumor shedding cells that directly returned to the primary), which they dubbed `primary seeding', was many orders of magnitude less likely than `secondary seeding', the process by which cells from the primary metastasize to a secondary location, grow and then re-shed progeny into the vasculature which then return to the primary.  This distinction, while difficult or currently impossible to measure in the clinic, is of chief importance, as it suggests that there are levels of detail about extant disease that are not captured in the previous models.  Specifically that the direct organ-organ `transitions' that were suggested by Newton et al. \cite{Newton:2013dg} could instead be meta-phenomena reflecting more than one transition, meaning that information could be missed concerning the location of metastatic colonies.

While we learn about the population-level propensity and temporal dynamics of spread from the models of Chen \cite{Chen:2009hz} and Newton \cite{Newton:2012bs,Newton:2013dg}, what is lacking is a framework by which these models could be applied to an individual patient. In order to make these models applicable to individual patients, and not just more accurate statements about population level data, we have to be able to tie them to clinical measurements.  While the model of Scott and colleagues is based on a highly heterogeneous selection of experiments \citep{Scott:2013hs}, the underlying framework is one that can be utilized in a patient-specific manner, a non-genetic application of the concept of `personalized medicine'.  Specifically, measurements of CTCs could be taken from the individual compartments (arterial, venous and portal venous, respectively, red, blue and purple in Figure \ref{fig:filterflow}) and used to infer the existence of subclinical metastatic disease.  This information would provide a better understanding of the overall tumor burden and would allow for clinical trials to test the utility of organ directed therapy or localized therapies, depending on the patient specific clinical data.

\section*{The way forward: communication and iterative multi-disciplinary science}

Mathematical models have several roles to play in the clinical and biological sciences.  The models presented in this review have highlighted a disparate set of these roles, including generation of novel hypotheses, explanation of phenomena which could not be described with existing, `cartoon' models, and prediction of patterns of spread.  We have specifically discussed a number of models of the metastatic process which lend insight to several different aspects of the process.  These insights include: the dynamics of the emergence of metastatic potential, the distribution of size of metastases through time, the possible mechanisms responsible for tumor dormancy, the patterns of spread of primary tumors, and possible mechanisms driving these patterns.

These theoretical models never stand alone in the scientific process, but they do represent an underutilized tool in the biological sciences, and particularly in the study of metastasis.  Metastasis remains the most important, lethal, and enigmatic part of cancer, and while we have been `waging war' against this disease for nearly 50 years, our progress has been limited.  Indeed, the limited scope of this review, which covers all of the theoretical work, to our knowledge, in this critical component of cancer progression, highlights the dire need for more work in this area.  More and more we are finding that scientists working alone are not able to make as much progress as they could working together - and indeed this has been the case in metastasis.  Going forward, scientists from disparate fields, including the mathematical/theoretical disciplines, must open and foster dialogues between one another, for if we aim to understand, and therefore interrupt, this complex and non-linear process, we have to work integrate and work together \cite{Anderson:2008to}.

\section*{Acknowledgment}

The authors would like to thank Katya Kadyshevskaya at the Scripps Institute for help in preparing Figure 6. JGS would like to thank the NIH Loan Repayment Program for support. AGF is funded by the EPSRC and Microsoft Research, Cambridge through grant EP/I017909/1.  PG, DB, ARAA and JGS gratefully acknowledge funding from the NCI Integrative Cancer
Biology Program (ICBP) grant U54 CA113007 and the Physical Sciences in Oncology Centers U54 CA143970.

\bibliography{metsreviewbib}{}
\bibliographystyle{unsrt}

\end{document}